\documentclass{jfm}
\usepackage{graphicx}
\usepackage{url}
\usepackage{xcolor}

\usepackage{lineno}

\shorttitle{Irreversibility of suspensions in highly confined duct flow}
\shortauthor{J. T. Antolik, A. Howard, F. Vereda, N. Ionkin, M. Maxey and D. M. Harris}
 
\title{{\color{black}Hydrodynamic irreversibility of non-Brownian suspensions in highly confined duct flow}}

\author{John T. Antolik\aff{1},    Amanda Howard\aff{2},
    Fernando Vereda\aff{3},
    Nikolay Ionkin\aff{1},
    Martin Maxey\aff{4}
    \and Daniel M. Harris\aff{1}\corresp{\email{daniel\_harris3@brown.edu}}}
    
\affiliation{
 \aff{1}
 School of Engineering, Brown University, Providence, Rhode Island 02912, USA
    \aff{2}Pacific Northwest National Laboratory, Richland, WA, 99354, USA
    \aff{3}Applied Physics Department, Faculty of Sciences,
University of Granada, Granada, Spain
\aff{4}Division of Applied Mathematics, Brown University, Providence, Rhode Island 02912, USA}
    
\begin{document}

\maketitle

\begin{abstract}
The irreversible behavior of a highly confined non-Brownian suspension of spherical particles at low Reynolds number in a Newtonian fluid is studied experimentally and numerically. 
In experiment, the suspension is confined in a thin rectangular channel that prevents complete particle overlap in the narrow dimension and subjected to an oscillatory pressure-driven flow.  In the small cross-sectional dimension particles rapidly separate to the walls, whereas in the large dimension  features reminiscent of shear-induced migration in bulk suspensions are recovered.  Furthermore, as a consequence of the channel geometry and the development and application of a single-camera particle tracking method, three-dimensional particle trajectories are obtained that allow us to directly associate relative particle proximity with the observed migration. Companion simulations of a steadily flowing suspension highly confined between parallel plates are conducted using the Force Coupling Method {\color{black} which also show rapid migration to the walls as well as other} salient features observed in the experiment. {\color{black} While we consider relatively low volume fractions compared to most prior work in the area, we nevertheless observe significant and rapid migration which we attribute to the high degree of confinement.}
\end{abstract}


\section{Introduction}

Suspension flows of spherical particles in a Newtonian fluid pertain to a wide range of applications, including biomedical, industrial, and geographical. Such flows represent generally chaotic dynamical systems \citep{drazer2002}, with the particles undergoing irreversible migration from an initial configuration to a final, steady state configuration. {\color{black} Mechanisms for particle or droplet migration across streamlines may be exploited for passive sorting in typical duct-like microfluidic channels \citep{marnoto2023application}.}

The role of particle contacts and surface roughness in non-Brownian suspensions is a topic of current interest as it relates to the particles' irreversible migration {\color{black}at low Reynolds} in a shear flow. Particle surface roughness results in an inherent irreversibility in the particles' relative motion as the particles are displaced across streamlines, and also modifies the pair distribution function between particles \citep{rampall1997, blanc2011, ingber2006}. It has been shown that neither the long-range hydrodynamic interactions \citep{metzger2010} nor the lubrication forces \citep{metzger2013} contribute significantly to the irreversibility. However, the particle surface roughness plays a key role in such dynamics \citep{zhang2023effect,pham2015,corte2008}, {\color{black} and can also directly impact the bulk rheological properties of non-Brownian suspensions \citep{more2020constitutive}.}
\cite{pine05} discovered that there is a critical strain amplitude for a given volume fraction below which the particle locations are reversible in oscillating flow. Above this critical strain amplitude, chaos sets in.  {\color{black} This critical strain amplitude is quite sensitive to the volume fraction, with irreversible behavior being more difficult to achieve with low volume fractions.} \cite{pham2016} found that smoother particles lead to a higher critical strain amplitude and predicted the critical strain as a function of the particle surface roughness, {\color{black}although very recent work has shown the role of particle roughness on irreversibility to be more subtle \citep{zhang2023effect}}. The critical strain amplitude has been further explored in flows with clouds of particles in an oscillating shear flow. The clouds are shown to extend until the volume fraction is less than the critical volume fraction for irreversibilty at a given strain amplitude \citep{metzger2012, howard2018}. {\color{black} The vast majority of prior experimental and numerical works have focused only on the simplest possible flow geometries, either flow between parallel plates or circular pipe flow, and typically in scenarios where the particles are much smaller than the confining geometry.  Taking a first step towards extending our understanding of particle migration behaviors and irreversibility to more general geometries with non-uniform shear in multiple directions and in scenarios where confinement effects are important are the primary goals of this work.}

Information about suspension flows comes from simulations and experiments. Simulations are costly, and quantities such as the particle surface roughness are represented by approximations such as an interparticle contact barrier. While simulations have been shown to accurately capture experiments \citep{maxey2017}, there is still a need for highly resolved experiments that can track particle locations. A number of experiments have been performed \citep{snook15, guasto2010, pham2016} that use a laser sheet to image a slice of the suspension. This technique requires the suspending fluid to be index-matched to the particles and complete trajectories are unavailable because particles are lost from view once they leave the laser plane. Laser-Doppler velocimetry (LDV) has also been used to achieve high-resolution measurements of particle velocity and concentration \citep{lyon98a, lyon98b} but the one-dimensional nature of LDV prohibits simultaneous capture of the entire flow. Magnetic resonance imaging (MRI) can provide 3D measurements of suspensions which are optically opaque and hence is well suited to suspensions in porous media \citep{mirbod2023}. However, individual particles cannot be tracked due to the relatively low resolution of MRI.  {\color{black} Single-camera refraction-based methods have been previously developed and applied to multiphase flows in the reconstruction of the three-dimensional shape of fluid interfaces \citep{moisy2009synthetic,kilbride2023pattern}; however to the best of our knowledge, similar ideas have not previously been adapted to suspension flows.} {\color{black} Although high-resolution 2D trajectory measurements have been performed previously to study irreversible dynamics in very small numbers of particles \citep{pham2015}, we believe our work is the first to provide such experimental data in a 3D random suspension of many particles, directly enabled by our new tracking technique.}

In this work, we {\color{black} experimentally} study the migration of {\color{black} spherical} particles suspended in a viscous fluid subjected to an oscillatory pressure driven {\color{black} duct} flow.   Extended 3D particle trajectories are resolved with a single camera {\color{black} by implementing a} refraction-based imaging technique capable of capturing particle motion in the out-of-plane direction. {\color{black} Although relatively low volume fractions are considered in this work, we nevertheless observe significant migration in both dimensions of the cross-section.} We report measurements of the migration dynamics {\color{black} and steady-state particle distributions} for different packing fractions and strain amplitudes, and {\color{black} quantitatively associate particle proximity during each cycle with the measured irreversibility.}  {\color{black} Our results are also directly compared to new numerical predictions of a highly confined suspension steadily flowing between parallel plates.}

\vspace{-3mm}

\section{Experiments}

\begin{figure}
    \centerline{\includegraphics[width=\textwidth]{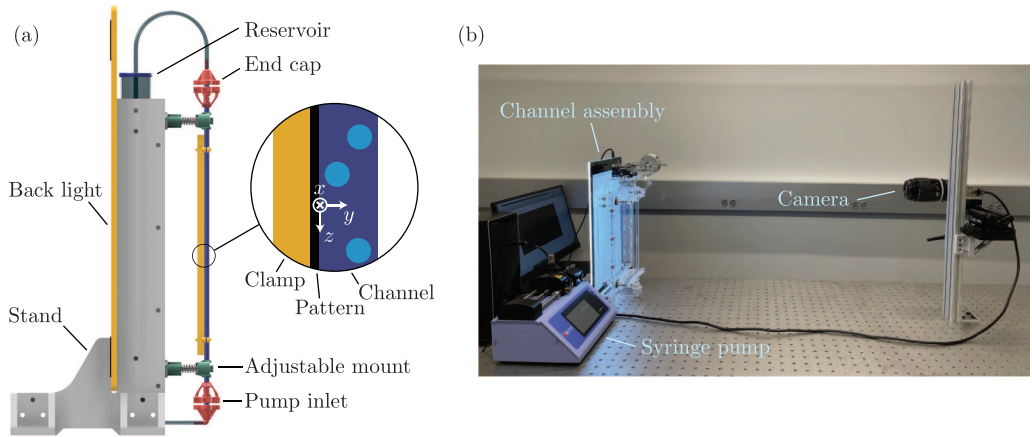}}
    \caption{(a) A diagram of the channel assembly, along with a magnified view of the rectangular glass channel and speckle pattern positioning. Channel cross-section is 3 mm ($y$-direction) $\times$ 9 mm ($x$-direction). (b) A photograph of the complete experimental setup.}
    \label{fig:setup}
\end{figure}

\begin{table}
    \centering
    \caption{Relevant parameters and their range of values in our experimental study.}
    \begin{tabular}{lccc}
        \textbf{Parameter}              & \textbf{Symbol}       & \textbf{Definition}           & \textbf{Value}                \\
        Channel width                   & $w$                   & --                            & 9 mm                          \\
        Channel height                  & $h$                   & --                            & 3 mm                          \\
        {\color{black}Channel length}                  & {\color{black}$l$}                   & --                            & {\color{black}300 mm}                          \\
        Particle diameter                 & $2a$                   & --                            & 1.598 $\pm$ 0.009 mm                        \\
        Particle RMS roughness          & --                    & --                            & 0.08 $\pm$ 0.03 $\mu$m                  \\
        Particle/fluid density          & $\rho$                & --                            & 1.179 $\pm$ 0.003 g/mL                      \\
        Fluid kinematic viscosity                 & $\mu$                 & --                            & 36.3 $\pm$ 1.8 cP                       \\
        Strain amplitude                & $\gamma$              & $QT/w^2h=\langle u_z \rangle T /w$                            & 1 - 6                         \\
        Accumulated strain              & $\gamma_a$            & $2Qt/w^2h=2 \langle u_z \rangle t /w$                            & 550 - 1100                    \\
        {\color{black}Number of particles} & {\color{black}$n$}                 &        --                                         & {\color{black}273 - 546} \\
        {\color{black}Nominal} {\color{black}bulk area fraction}            & {\color{black}$\phi_A'$}              & {\color{black}$\pi a^2 n / (wl)$}                            & {\color{black}0.2 - 0.4}                     \\ 
        {\color{black}Nominal} bulk volume fraction            & {\color{black}$\phi_B'$}             & {\color{black}$4/3\pi a^3 n / (wlh)$}                            & {\color{black}0.07 - 0.14}     \\
        {\color{black}Actual bulk volume fraction}             & {\color{black}$\phi_B$}              & --               & {\color{black}0.06 - 0.12} \\
        Volumetric flow rate            & $Q$                   & $\langle u_z \rangle w h$                           & 1 $\pm$ 0.005 mL/min                      \\
        Particle diffusion coefficient  & $D$                   & $kT_\textrm{abs}/(6\pi\mu a)$                            & $7.4 \times 10^{-18}$ m$^2$/s \\
        Particle Reynolds number                 & $\Rey$                & $\rho Q a / (w h \mu)$        & $1.6 \times 10^{-2}$          \\   
        Peclet number                   & $\Pen$                & $a Q / (w h D)$               & $8.8 \times 10^{11}$             
    \end{tabular}
\label{tab:parameters}
\end{table}

The particles used in our experiments are optically clear acrylic (PMMA) spheres with diameter 1.598 $\pm$ 0.009 mm and root mean square surface roughness 0.08 $\pm$ 0.03 $\mu$m.  They are suspended in a water-glycerol mixture with Newtonian viscosity 36.3 $\pm$ 1.8 cP \citep{cheng08} filling a 300 mm long Borosilicate glass test channel (VitroCom R0309, rectangular internal cross section 9 $\times$ 3 mm). {\color{black}The mixture ratio is chosen such that the particles are neutrally buoyant, each with a density of 1.179 $\pm$ 0.003 g/mL.} The 3 mm channel height is just less than two particle diameters,  preventing complete particle overlap.  {\color{black} The 9 mm channel width was selected to be small enough so that a significant shear gradient was present across the entire $x$-direction, and wide enough to allow particles to move a few diameters away from the side walls.} {\color{black}Experiments performed by \citet{guasto2010} use the same channel aspect ratio, though in their case the particle size to channel depth ratio is far smaller and only a single central plane is imaged, with the intent of approximating a two-dimensional flow.}
The complete experimental setup is shown in figure \ref{fig:setup} {\color{black} and dimensional and non-dimensional parameters summarized in table \ref{tab:parameters}}.
The glass test channel is oriented vertically and attached to a laser cut stand by screw-adjustable mounts. 
3D-printed end caps with flow orifices 
ensure the particles remain in the channel and adapt the ends of the channel to hoses. The upper hose goes to a reservoir of suspending fluid which is open to the atmosphere and the bottom hose attaches to a syringe pump (Harvard Apparatus Pump 11 Elite). 

A back light panel (3W LED tracing pad) illuminates the test section. We generate speckle patterns using the Rosta algorithm implementation (dot size 1, density 0.3 and zero smoothing) in the $\mu$DIC Python library \citep{olufsen19} which are then ink-jet printed with 1200 x 2400 DPI resolution on transparency film. 
The pattern is fastened to the channel with a transparent acrylic clamping block.
An Allied Vision Mako 503B monochrome CMOS camera with 105 mm micro Nikkor lens is used for imaging the suspension at a 1 Hz frame rate and 2592 $\times$ 610 resolution. The camera is placed at a working distance of approximately 1 m, resulting in a 40 mm by 10 mm imaged section. Hence lens distortion is minimized and the incident rays from the camera can be assumed to be parallel. The outlets of the channel are 130 mm away from the imaged section so entry length effects can be neglected. 


We prepare particle suspensions with {\color{black}nominal} bulk particle volume fractions ${\color{black}\phi_B'} =$ {\color{black}0.07, 0.11 and 0.14}. {\color{black}The corresponding {\color{black}nominal} bulk area fractions {\color{black}$\phi_A'$} -- defined as the area of the experimental image that would be filled with particles assuming no overlap -- are 0.2, 0.3 and 0.4. Similar bulk area fraction values have been previously used in simulations of a monolayer suspension \citep{nott1994pressure}.}
A magnetic 3D-printed plow-shaped particle pusher is used to manually set the initial particle distribution in the channel by guiding it with a handheld magnet outside the channel. The geometry of the pusher is such that the particles are driven to the channel walls in the $x$ direction and the channel center ($y=h/2$) in the $y$ direction. 
However, we observe that the particles quickly migrate towards the walls along the $y$ direction once the experiment starts. {\color{black}Since the manual particle arrangement process produces some heterogeneity in the particle distribution, the measured bulk volume fraction $\phi_B$ in the field of view (averaged over all frames of the experiment) is slightly different than the nominal $\phi_B'$, and takes values ranging from 0.06 to 0.12.}

During the experiment, the syringe pump generates oscillatory flow in a square wave, prescribing the volumetric flow rate $Q$ to be
\begin{equation}
    Q(t) = \frac{\gamma w^2 h}{T} \frac{\sin(2 \pi t / T)}{|\sin(2 \pi t / T)|}, 
\end{equation}
where $T$ is the flow oscillation period. The strain amplitude $\gamma$ is the average displacement of a fluid parcel over a half cycle $T/2$ divided by the channel half width $w/2$. We perform experiments with $\gamma=6$ up to an accumulated strain of 550 (46 oscillation cycles) and with $\gamma=1$ to an accumulated strain of 1100 (550 oscillation cycles). The period $T$ is chosen to maintain a particle Reynolds number of 0.016.
Initial experiments performed at a Reynolds number of 0.008 produced indistinguishable suspension dynamics.
Experiments were also performed with a single particle in the channel showing fully reversible motion in 3D, regardless of its initial position.

\begin{figure}
    \centerline{\includegraphics[width=\textwidth]{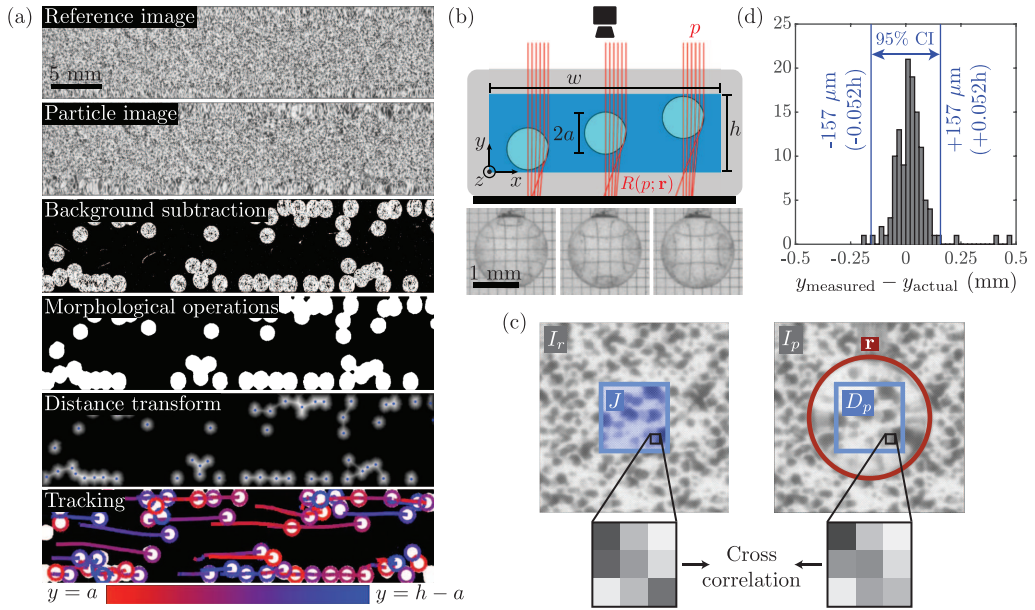}}
    \caption{(a) Example images of the channel test section at each step of the processing procedure. The tails in the ``tracking" panel illustrate how each particle's position has evolved from the initial state in the previous panels, with the coloring indicating the $y$ position over time. (b) Cross-sectional diagram of the channel labelled with relevant dimensions. Incident rays from the camera at $p$ are propagated based on the refraction model to $R(p;\mathbf{r})$ for particles at different heights. Experimental images of particles over a regular 200 $\mu$m grid pattern at the different heights shown below. (c) The reference image $I_r$ is transformed based on the refraction model and the guessed particle position $\mathbf{r}$. The result in $J$ is compared to the corresponding region $D_p$ in the particle image $I_p$ using the cross correlation. (d) Histogram of particle height measurement errors from 120 data points taken at 12 evenly spaced known positions in $y$.} 
    \label{fig:processing}
\end{figure}

Prior to each experiment, a particle-free reference image of the speckle pattern is taken. Then, throughout the experiment, a series of photographs of the test section with particles are captured. Figure \ref{fig:processing}(a) outlines the image processing steps which include background subtraction \citep{zivkovic04}, filtering with morphological operations, and determining 2D particle positions with the Euclidean distance transform and h-maxima filter \citep{vincent93}. The out-of-plane $y$-component of each particle's position may be inferred by comparing the reference and particle images since the speckle pattern in the particle image will be distorted depending on the particles' positions. Figure \ref{fig:processing}(b) illustrates the working principle of this particle tracking technique. The incident rays from the camera refract at the fluid-particle interfaces and focus as they reach the pattern attached to the outside of the channel. As seen in the photographs corresponding to each particle height, increasing the $y$ component of the particle position has the approximate effect of increasing the magnification of the pattern. By applying Snell's law at each optical interface in this axisymmetric geometry \citep{hecht2012}, an analytical expression may be constructed which predicts the observed distortion of the reference pattern through the transparent particle. For the case of partially overlapping particles, the distortion is determined numerically by recursively propagating each incident ray from the camera through the cluster of particles until it reaches the speckle pattern. The particle's position is then determined by solving the inverse problem illustrated in \ref{fig:processing}(c): given a reference image $I_r$ of the pattern and the resulting particle image $I_p$, we find a position $\mathbf{r}$ for the particle that reproduces the observed distortion by maximizing the cross correlation between regions $J$ and $D_p$. We use the Nelder-Mead algorithm \citep{nelder65} in the NLopt library \citep{nlopt} to optimize the cross correlation. The OpenCV library \citep{opencv} for C++ is used to implement the image transformations and cross correlations. {\color{black}The tracking method requires calibration in order to precisely determine the refraction indices of the particles, fluid and channel walls. A calibration device was constructed by attaching a particle on a thin wire to a high precision translation stage. The particle may then be moved to known $y$ positions across the height of the channel while capturing images of the distorted speckle pattern. Using the algorithms described, the particle positions are measured from each image and the refraction indices are adjusted in order to minimize the sum of squared residuals between the actual particle position and the measured position. After calibration, the translation stage was again used to independently determine the accuracy of the tracking method.} 
Using a number of different particles at various locations in the field of view, the comparison between the measured height and the actual height reveals excellent agreement (figure \ref{fig:processing}(d)). Over the full measurement range, the particle's $y$ position can be determined to within 157 $\mu$m or 5.23\% of the channel height with 95\% confidence. The final processing step is to link the 3D particle positions into trajectories using Crocker and Grier's algorithm \citep{crocker96} in the open-source library TrackPy \citep{trackpy}  so that each particle's identity is known throughout multiple frames, with the result shown in the ``tracking" panel of figure \ref{fig:processing}(a).

{\color{black}
\section{Simulations}
To compare with the experiments, we also completed a limited set of simulations using the Force Coupling Method (FCM) \citep{Yeo2010, Yeo2011}. Due to computational limitations, the simulations are completed with a single set of walls located at $y = 0$ and $y=4a$ representing a very narrow channel {\color{black} (corresponding to a channel height of $h=4a$)}, and are periodic in the streamwise ($z$) and spanwise ($x$) directions. The simulations are scaled by the particle radius $a=1.$
To set up the simulations, the particles are seeded randomly in the channel to reach the proper area fraction. The particles are first seeded with radius $0.85a$, and then are ``inflated'' to reach radius $a = 1$ with a molecular dynamics simulation. In doing this work, we experimented with different methods of seeding the particles to match the experimental results. With simulations, it is possible to seed the particles in a perfect monolayer, located at the channel half height {\color{black}$y =2a$}. However, this does not match the experiments, because the simulation particles are not subject to noise in their location, and therefore would remain in a monolayer for an unphysically long time. Instead, we seed the particles in a ``pseudo-monolayer,'' or a noisy monolayer. The particles are seeded with average $y$ location at {\color{black}$y=2a$}, and a standard deviation of $0.05a$. The minimum $y$ value for $\phi_B = 0.14$  is $1.833a$, and the maximum $y$ value is $2.221a$ at the beginning of the simulation. This way, the particles begin close to the center of the channel, but particle interactions allow for the particles to migrate in the cross-stream ($y$) direction. The channel width is $w = 30a$ and the channel length is $l = 60a$, which results in a total of $115$ particles for $\phi_B = 0.07$, $172$ particles for $\phi_B = 0.11$, and $230$ particles for $\phi_B = 0.14$.

The flow is driven by a pressure gradient $\partial p / \partial z = 0.5$. Because the strain amplitude in the experiments are calculated using the $x-$direction length scales, we calculate the accumulated strain as $\gamma_a = 2\langle u_z \rangle t / (11 a)$, where the scaling factor of $11$ is chosen to approximate the experimental strain amplitude scaling factor of $w$.

The simulations use a time step of $\Delta t = 0.005$. To approximate the particle surface roughness, we use a short range contact barrier that disrupts the symmetry in particle interactions and computationally prevents the particles from overlapping. The contact force between particles $\alpha$ and $\beta$, with centers $\mathbf{r}^{\alpha}$ and $\mathbf{r}^{\beta}$, acts along the line of centers of the particles and is given by
\begin{equation}
\mathbf{F}^{\alpha \beta}_P = \left\{ \begin{array}{ll}  
\displaystyle -6\pi \mu a V_{ref} \left( \frac{R_{ref}^2-|\mathbf{r}|^2}{R_{ref}^2-4  a^2} \right)^6 \frac{\mathbf{r}}{|\mathbf{r}|} 
  \quad \mbox{if\ } \quad |\mathbf{r}| < R_{ref},\\[8pt]
\displaystyle  \quad 0
  \quad \mbox{otherwise\ }
 \end{array}\right.
  \label{eq:ForceBarrier}
\end{equation}
in which $\mathbf{r} = \mathbf{r}^\beta - \mathbf{r}^\alpha$ and $R_{ref}$ is the cut-off distance, $R_{ref} = 2.01a$ \citep{Yeo2010, Yeo2011}. The relationship between the contact force reference values and the pressure gradient driving the flow sets the minimum separation between particles. $V_{ref}=200$ is a constant chosen to so that the minimum gap between two particles is approximately $0.005a$. Particles separated by a distance of less than $0.01a$ are considered to be in contact, so $R_{ref}= 2.01a$. }
\textcolor{black}{
The contact force between a particle $\alpha$ with center $\mathbf{r}^{\alpha}$ and a wall is given by: 
\begin{equation}
\mathbf{F}^{\alpha, wall}_P = \left\{ \begin{array}{ll}  
\displaystyle -6\pi \mu a V_{ref, wall} \left( \frac{R_{ref, wall}^2-|\mathbf{r}|^2}{R_{ref, wall}^2-a^2} \right)^6 \frac{\mathbf{r}}{|\mathbf{r}|} 
  \quad \mbox{if\ } \quad |\mathbf{r}| < R_{ref, wall}\\[8pt]
\displaystyle  \quad 0
  \quad \mbox{otherwise\ }
 \end{array}\right.
  \label{eq:ForceBarrier_wall}
\end{equation}
where $\mathbf{r}$ is the vector between $\mathbf{r}^{\alpha}$ and the top or bottom wall. We take $R_{ref, wall} = 1.05a$ and $V_{ref, wall} = 200.$ 
}

\section{Results}

\subsection{{\color{black}Migration to the walls: cross-stream $y$-direction}}

\begin{figure}
    \centerline{\includegraphics[width=\textwidth]{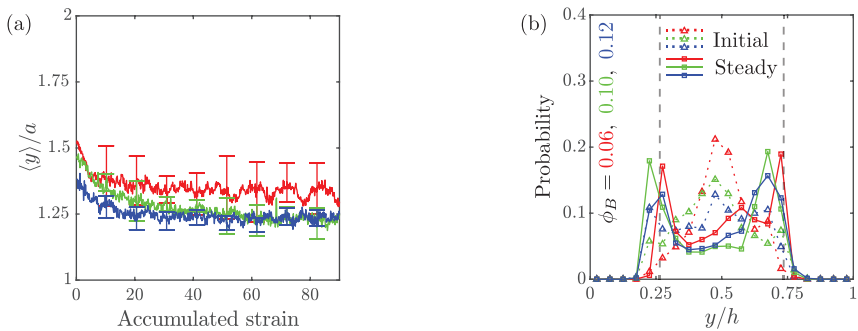}}
    \caption{{\color{black}Experimental results for migration in the cross-stream ($y$) direction. (a) Average distance between a particle center and the nearest wall versus accumulated strain for experiments with $\gamma=6$ and $\phi_B =$ {\color{black}0.06} (red), {\color{black}0.10} (green) and {\color{black}0.12} (blue). The error bars show the standard deviation between 5 trials. (b) Initial and steady state particle position distributions over the height of the channel show that, in the narrow channel dimension, the particles prefer the walls at steady state. The initial distribution is averaged over the first 20 frames ($\gamma_a < 2.74$) and the steady profile over frames 601 to 656 ($\gamma_a = [82, 90]$). The vertical dashed lines indicate a distance of one particle radius from the wall.}}
    \label{fig:narrow_migration_exp}
\end{figure}

\begin{figure}
    \centerline{\includegraphics[width=\textwidth]{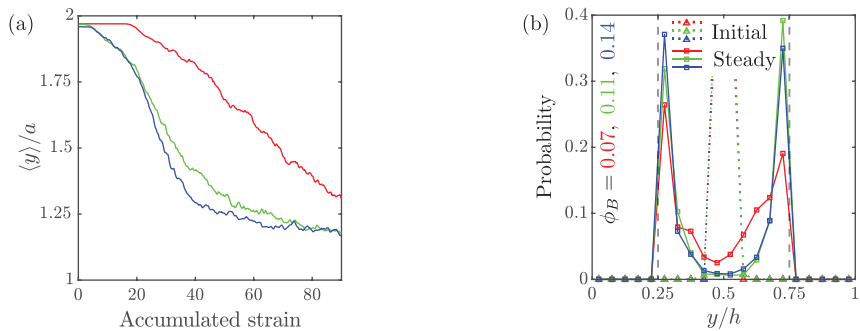}}
    \caption{{\color{black} Simulation results  for migration in the cross-stream ($y$) direction. (a) Average distance between a particle center and the nearest wall versus accumulated strain for simulations with steady flow and $\phi_B =$ 0.07 (red), 0.11 (green) and 0.14 (blue). The particles begin in a ``pseudo-monolayer'' but quickly migrate to the walls in the narrow channel dimension. (b) Initial and steady state particle position distributions over the height of the channel. The initial distribution is averaged over the first 20 frames ($\gamma_a < 3$) and the steady profile over frames 550 to 600 ($\gamma_a = [82, 90]$). The vertical dashed lines indicate a distance of one particle radius from the wall.}}
    \label{fig:narrow_migration_sim}
\end{figure}

{\color{black}Although we bias the particles to the channel center ($y=h/2$) in the narrow dimension at the start of the experiment, we observe that the particles rapidly migrate towards the walls once the suspension is subjected to a periodic strain. The rate of this migration is captured in figure \ref{fig:narrow_migration_exp}(a) which shows the average distance between a particle center and the nearest wall as a function of accumulated strain $\gamma_a$. On a timescale which is approximately an order of magnitude faster than the less confined $x$-direction evolution we will present in section \ref{x_migration}, the particles reach a configuration in which the average particle is about 1.25 radii away from the channel wall. The corresponding FCM simulations produce a similar result in figure \ref{fig:narrow_migration_sim}(a) where the migration to the walls is even more apparent due to the near-monolayer initial condition. The rate of migration is enhanced for higher $\phi_B$. For the simulations, the final average $y$-locations for the two highest values of $\phi_B$ are about $1.25a$, the same as for the experiments.  The initial and steady state particle position distributions in the experiments and simulations are presented in figures \ref{fig:narrow_migration_exp}(b) and \ref{fig:narrow_migration_sim}(b), where we plot the probability that a given particle's center will lie in one of 20 evenly spaced bins distributed over the channel height. Both the experimental and simulation plots show an initial distribution that favors the channel center due to the imposed particle configuration. However, the particles strongly prefer the walls in the steady state distribution. {\color{black}The data points which appear closer than one particle radius to the wall in the experiments are a consequence of the $y$-direction tracking uncertainty, which is quantified in figure \ref{fig:processing}(d).} {\color{black}The experimental measurements of \cite{snook15} in circular pipe flow show the presence of local peaks in the radial concentration distribution at the wall which they attribute to a particle wall-layering effect. This becomes more pronounced in their data for the smaller pipe radius considered, an effect they attribute to the higher confinement.  In the case of extreme $y$-direction confinement studied here, such wall layers dominate and ultimately overwhelm the final concentration distribution.  As also noted by \cite{snook15}, continuum models such as the Suspension Balance Model are of course unable to capture such discrete particle effects.}

It is possible to understand this migration to the walls by considering two-body interactions between the particles. If two particles come into contact, the force between the particles acts along the line of centers of the particles. Unless the particles have identical $y-$locations, the force will have a component that displaces each particle towards the walls. After many strain units and particle interactions, the particles will migrate to be close to the walls. Once a particle enters the wall layer, that is, it is on top of the wall, it is very difficult for the particle to exit because any additional interaction will force the particle closer to the wall.}

\subsection{{\color{black}Migration to the center: spanwise $x$-direction}} \label{x_migration}

\begin{figure}
    \centerline{\includegraphics[width=\textwidth]{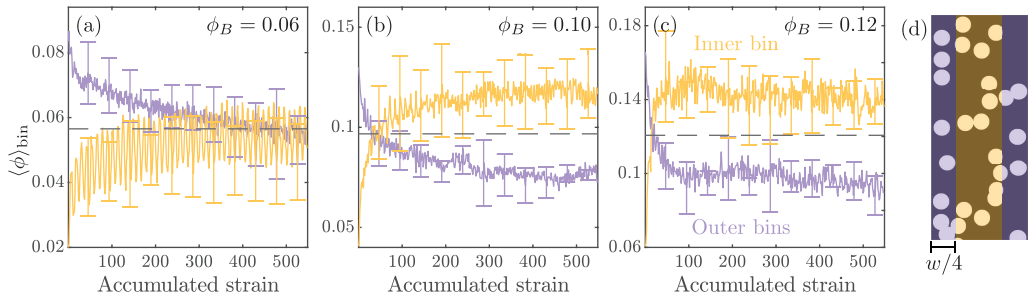}}
    \caption{Experimental results for migration in the spanwise ($x$) direction.  The evolution of particle concentration in the inner and outer bins is shown for (a) $\phi_B={\color{black}0.06}$, (b) $\phi_B={\color{black}0.10}$ and (c) $\phi_B={\color{black}0.12}$; $\gamma = 6$. The dashed line indicates the {\color{black}measured} bulk {\color{black}volume} fraction {\color{black}$\phi_B$} in the field of view averaged over the experiment. (d) The outer bins are one quarter of the channel width and the concentrations are calculated from the {\color{black}reconstructed 3D particle positions as the volume in the bin filled with particles divided by the total bin volume.}}
    \label{fig:bin_evolution}
\end{figure}

Examining the evolution of the local particle {\color{black}volume} fraction $\phi$ as a function of accumulated strain $\gamma_a$ gives insight into the dynamics of the migration process and its dependence on bulk packing fraction. The average particle concentrations in the center and outer quarter-width bins when $\gamma=6$ are plotted against the accumulated strain in figure \ref{fig:bin_evolution}. {\color{black}The bin concentration $\langle \phi \rangle_\mathrm{bin}$ is computed based on the 3D particle positions as the volume in the bin filled with particles divided by the total bin volume.} {\color{black}When a particle intersects the $x=w/4$ or $x=3w/4$ planes, its volume is partitioned accordingly between the inner and outer bins.} The results in these and all subsequent {\color{black}experimental} plots are averaged over at least 5 trials, with error bars representing one standard deviation. Due to the initial configuration of the particles, the initial outer bin concentration exceeds the bulk packing fraction for all experiments. However, for the $\phi_B={\color{black}0.10}$ and $\phi_B={\color{black}0.12}$ experiments, the inner and outer bin concentrations eventually cross and, at high accumulated strain, the inner bin concentration exceeds the bulk packing due to particle migration. Migration proceeds at an increased rate for $\phi_B={\color{black}0.12}$. For the low packing fraction experiments with $\phi_B={\color{black}0.06}$, the bin concentrations do not cross the average channel concentration up to an accumulated strain of 550, indicating a sharp reduction in the rate of migration. The oscillations in the inner bin concentration in figure \ref{fig:bin_evolution}(a) match the period of the flow and occur due to small variations in the packing fraction along the length of the channel. {\color{black} ``Anomalous'' migration was not observed in the present experiments, but is predicted to occur for significantly smaller strain amplitudes than explored in the present work \citep{morris2001anomalous}.}

\begin{figure}
    \centerline{\includegraphics[width=\textwidth]{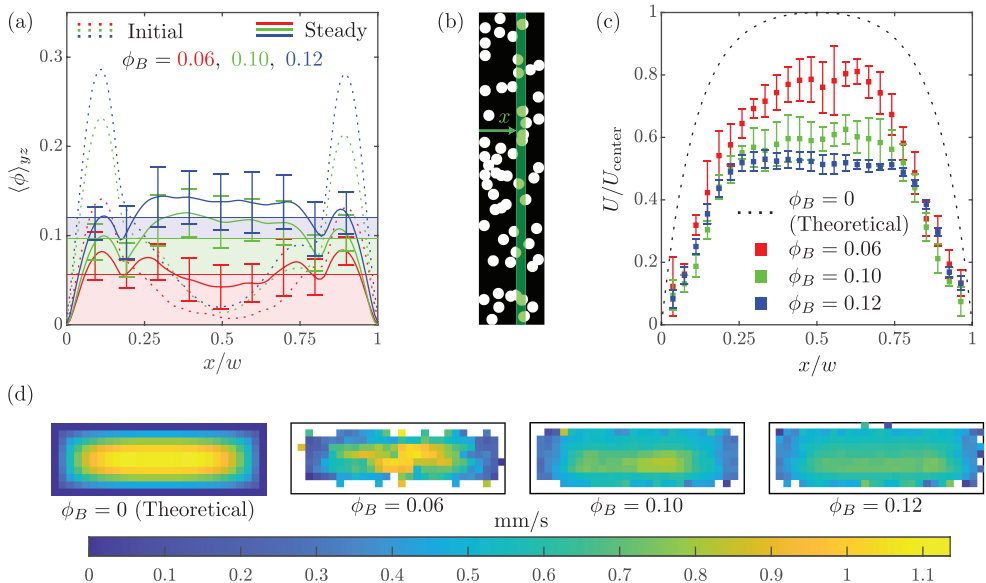}}
    \caption{(a) Initial and steady state concentration profiles for $\gamma=6$. Initial profiles were averaged over the first 20 frames ($\gamma_a < 2.74$) and steady profiles over frames 3500 to 3700 ($\gamma_a = [480, 508]$). {\color{black}The shaded regions indicate the measured bulk packing fraction $\phi_B$ in each experiment.} (b) A bin {\color{black}of width $w/100$} is swept over the {\color{black}reconstructed 3D particle positions} in $x$ to measure local {\color{black}volume} concentration {\color{black}$\langle \phi \rangle_{yz}$}. (c) Center-line ($y=h/2$) particle velocity profiles compared to theoretical Newtonian case. {\color{black}The curves are normalized by the theoretical velocity at the channel center $U_\mathrm{center} =$ 1.15 mm/s.} (d) Cross-sectional particle velocity profiles.}
    \label{fig:profiles}
\end{figure}

The initial and steady state particle concentration profiles over the channel width are shown in figure \ref{fig:profiles}(a) for experiments with $\gamma=6$. Since the experiments begin in a wall-loaded configuration, the initial concentration profiles contain two peaks that are approximately one particle radius away from the channel walls. {\color{black} Additional experiments were conducted for more homogeneous initial concentrations and showed very similar results (Appendix A).} For the $\phi_B={\color{black}0.10}$ and $\phi_B={\color{black}0.12}$ experiments, the steady state concentration profiles indicate significant migration towards the channel center since the bulk packing fraction is surpassed there. However, compared against results with a parabolic flow profile in \citet{snook15} and \citet{lyon98a}, the {\color{black} enhanced concentration} at the channel center is less pronounced, mirroring the flatter velocity profile in the current geometry. For $\phi_B={\color{black}0.06}$, the particle arrangement relaxes from its initial configuration suggesting some irreversible behavior but there is less preferential migration towards the center of the channel. {\color{black}The $x$-direction concentration profiles are constructed by computing the local $y$-and-$z$-averaged particle volume concentration $\langle \phi \rangle_{yz}$ in a bin which is swept over the width of the channel as shown in figure \ref{fig:profiles}(b). The bin width is $w/100$} {\color{black}and the volume concentration $\langle \phi \rangle_{yz}$ is computed as the volume of the bin filled with particles divided by the total bin volume.}  
For clarity, error bars are shown at discrete points rather than at every {\color{black}bin station}. {\color{black} While we do observe a net migration of particles to regions of lower shear in the channel center (in the $x$-direction), we note that due to the relatively large particle to channel size (i.e. high confinement) it is unlikely that our results would be well described by continuum models of similar behaviors such as the Suspension Balance Model \citep{morris1999curvilinear,guazzelli2018rheology}.}  {\color{black} In contrast to the present work, prior experimental work on shear-induced migration in less confined suspensions in circular pipes have suggested an absence of migration for suspensions around $\phi_B\approx 10 \%$ in both oscillatory \citep{snook15} and steady flows \citep{hampton1997migration}.  The fact that detectable migration towards the center is observed for the relatively low volume fractions in our system is thus likely a consequence of the high levels of confinement, which naturally increase the likelihood of particle contacts. Such contacts are known to be the primary source of irreversibility in these systems \citep{metzger2013}.}

\subsection{{\color{black}Migration rate}}
We quantify the rate of migration in our experiments by measuring the number of oscillations required for the particle concentration profile to reach steady state following a method similar to the one used in \cite{snook15}. As seen by comparing figures \ref{fig:narrow_migration_exp}(a) and \ref{fig:bin_evolution}, the time to steady state in our experiments is dominated by the migration in the $x$-direction. As such, in order to determine steady state, we compute $\langle \phi \rangle_{yz}$ using bins of width $w/20$ for each oscillation of the experiment, averaging over frames 22 to 76 of the oscillation. When a majority of the bins are within one standard deviation of the particle concentration over the remaining oscillations, the suspension is considered to have reached steady state. Steady state is reached in 29 $\pm$ 14, 11 $\pm$ 7 and 4 $\pm$ 1 oscillations for experiments with $\phi_B =$ {\color{black}0.06, 0.10, and 0.12}, respectively. {\color{black} The values following the means characterize the distribution in the time to steady state between trials.}  This result matches the observation in \cite{snook15} that the {\color{black} mean} accumulated strain to reach steady state scales approximately as $\phi_B^{-2}$. {\color{black} Despite the lower volume fractions considered in our work, the mean number of oscillations to steady state are comparable in magnitude to the numerical values reported by \cite{snook15}.  Furthermore their data indicates a decrease in time to steady state with both volume fraction and degree of confinement.  It is thus likely that the increased confinement in our system promotes more particle contacts than in an otherwise equivalent lesser confined scenario, which compensates for the reduced likelihood of contact associated with smaller volume fractions.}   

\subsection{{\color{black}Particle velocities}}
Enabled by the 3D particle tracking, figure \ref{fig:profiles}(d) shows the average magnitude of the particle velocity over the channel cross section. The experimental velocity profiles are time-averaged over a full experiment and the theoretical velocity is calculated for a pure Newtonian fluid in Stokes flow. The white cells in the experimental profiles are regions where no velocity measurement could be obtained since the particle centers must be at least one particle radius away from the channel walls.
For the $\phi_B={\color{black}0.06}$ case, the magnitude of the measured velocity profile is close to the theoretical profile but the particle velocities are reduced for the higher packing cases. The mid-line particle velocity profiles at $y=h/2$ are compared in figure \ref{fig:profiles}(c) against the theoretical center-line velocity. {\color{black}The experimental velocity profiles are constructed using particles whose centers are within $0.12h$ of the channel center.} The profile becomes increasingly blunt with higher $\phi_B$, consistent with results in a circular pipe \citep{snook15} and two-dimensional channel flow \citep{lyon98a,guasto2010}.

\begin{figure}
    \centerline{\includegraphics[width=\textwidth]{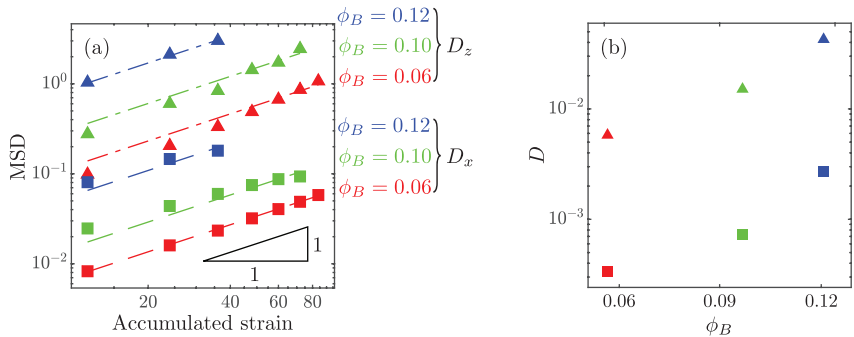}}
    \caption{(a) Mean squared particle displacements in $x$ and $z$ for $\gamma=6$ experiments.  The dashed lines show linear fits to the data used to extract the effective diffusivities via equation (\ref{eqn:D}). (b) Comparison of effective diffusivities $D_x$ (squares) and $D_z$ (triangles) at different bulk volume fractions.}
    \label{fig:diffusion}
\end{figure}

\begin{figure}
    \centerline{\includegraphics[width=\textwidth]{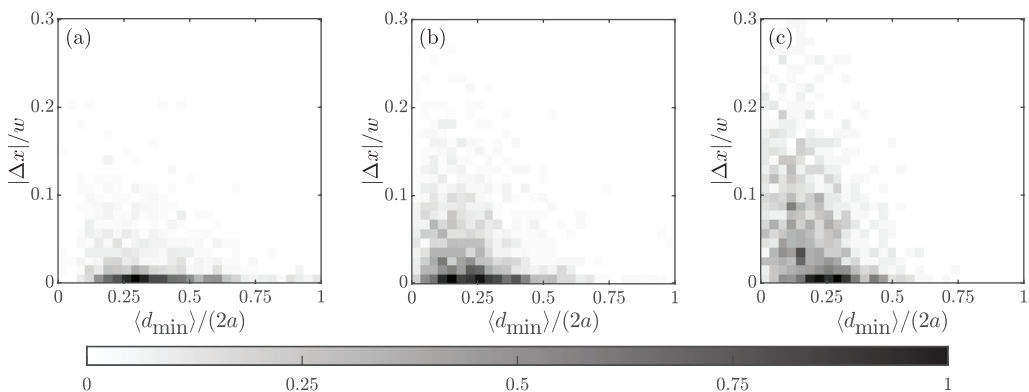}}
    \caption{{\color{black}Experimental} heat maps of {\color{black}$x$-}particle migration over one cycle versus average distance to nearest neighbor at $\gamma=6$ for (a) $\phi_B={\color{black}0.06}$, (b) $\phi_B={\color{black}0.10}$ and (c) $\phi_B={\color{black}0.12}$. The color map is normalized by the maximum bin count which is 56, 57 and 32, respectively.}
    \label{fig:heatmap}
\end{figure}

\begin{figure}
    \centerline{\includegraphics[width=\textwidth]{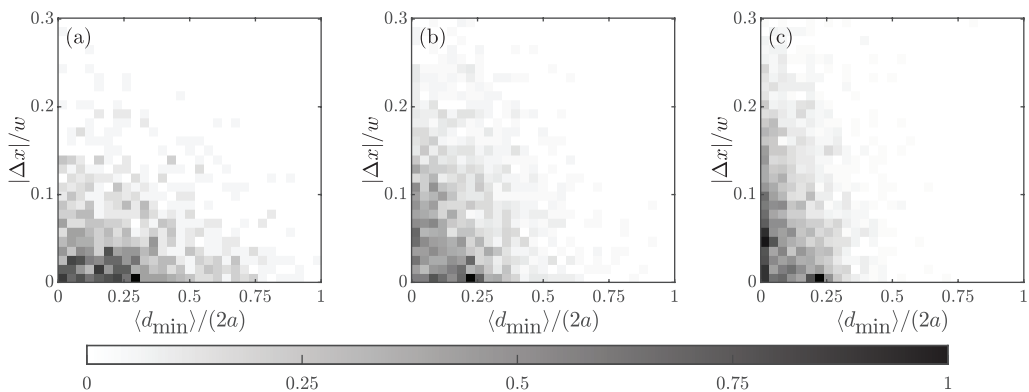}}
    \caption{{\color{black}Simulation heat maps of $x$-particle migration after 12 strain units versus average distance to nearest neighbor for (a) $\phi_B=0.07$, (b) $\phi_B=0.11$ and (c) $\phi_B=0.14$. The color map is normalized by the maximum bin count which is 22, 42 and 54, respectively.}}
    \label{fig:heatmap_simulation}
\end{figure}

\subsection{{\color{black}Particle irreversibility and interactions}}
The mean square particle displacements (MSD) are a quantitative measure of irreversibility in the particles' motion and thus allow more detailed comparison between the experiments. The non-dimensional effective particle diffusivity $D_x$ is defined following \citet{pine05} according to
\begin{equation}
    \langle (\Delta x /(2 a))^2 \rangle= 2 D_x \gamma_a\label{eqn:D}
\end{equation}
where $\langle (\Delta x /(2a))^2 \rangle$ is the non-dimensional MSD, $2a$ is the particle diameter, $\gamma_a$ is the total accumulated strain and $\Delta x$ is the displacement of the particle over one cycle. Given the three-dimensional nature of the flow geometry, effective diffusivities $D_y$ and $D_z$ for the other directions are defined in a similar manner. In order to ensure that the results are statistically significant, MSDs are computed up to the maximum accumulated strain for which there are at least 50 continuous particle trajectories (starting at any point in the experiment). Since higher bulk packing fraction increases the likelihood that a particle is lost during tracking, MSDs are calculated to a lower accumulated strain for higher $\phi_B$ experiments.  
The MSDs after an integral number of cycles in the $\gamma=6$ experiments are graphed against the accumulated strain in figure \ref{fig:diffusion}(a). The $y$-component is omitted as it does not exhibit diffusive behavior, presumably due to the high confinement. The least squares fits for the effective particle diffusivities are shown as well. For a given packing fraction, the effective diffusivity along the flow direction $D_z$ is significantly higher than the {\color{black}spanwise} diffusivity $D_x$. This result is consistent with findings in a circular Couette flow  where two {\color{black} or more} components of diffusivity were measured {\color{black}\citep{pine05, breedveld2001, breedveld2002}}. \citet{guasto2010} also report that the streamwise diffusivity is consistently higher than the spanwise diffusivity in a rectangular channel flow. Figure \ref{fig:diffusion}(b) compares the effective diffusivities as a function of $\phi_B$. For every value of $\phi_B$, the ratio between $D_x$ and $D_z$ is about 20 but their values increase approximately one order of magnitude as $\phi_B$ increases from {\color{black}0.06} to {\color{black}0.12}. The increase of effective diffusivity with bulk volume fraction is consistent with experimental results in the literature for other flow geometries \citep{pine05, guasto2010}.  {\color{black} We note that our experiments involve non-uniform strain and time-evolving non-uniform particle concentrations.  As such, the particle diffusivities will depend on space \citep{guasto2010} and time \--- dependencies that a single particle diffusivity value does not capture. {\color{black}Even in the case of a simple shear flow, it has been observed experimentally that the particle diffusivity can have distinct short and long-time values \citep{breedveld2001}.} Nevertheless, the average diffusivities computed here remain a reasonable quantitative proxy for the general level of chaos in the system \citep{drazer2002}, and allow for a simple comparison between our experimental conditions.  Direct quantitative comparison of these numerical values with simpler cases involving uniform strain and particle concentration such as \citet{pine05} should be made with care, however. With significantly more data, the spatial dependence and temporal evolution of the particle diffusivities could be faithfully resolved using our experimental setup.}

Since the 3D position of each particle in the experiment is known, interactions between particles and the resulting irreversibility may be directly examined. Figure \ref{fig:heatmap} shows the number of particles that migrate a given amount based on their proximity to their neighbors over a cycle. A particle's level of interaction is quantified {\color{black} along the abscissa} by {\color{black}$\langle d_{\mathrm{min}} \rangle/(2a)$} which represents the average {\color{black}3D} {\color{black}surface-to-surface} distance between a particle and its nearest neighbor over a cycle, normalized by the particle {\color{black}diameter}. The nearest neighbor is chosen at each frame in the cycle and the data is collected over the entire experiment. {\color{black}While other metrics for particle contact pressure such as the radial distribution function have been used in the literature \citep{rintoul96}, we find that our metric $\langle d_{\mathrm{min}} \rangle/(2a)$ is more strongly correlated to the irreversible particle displacements in our experiments.} {\color{black} The magnitude of the spanwise displacement over a single cycle $|\Delta x|$ is plotted on the ordinate, and is a measure of the irreversibility of particle motion during the cycle.  These plots demonstrate that} for all experiments, the particles with more interactions {\color{black}(smaller $\langle d_{\mathrm{min}} \rangle$)} have a greater tendency to migrate {\color{black}(larger $|\Delta x|)$}.  {\color{black}The presence of close interactions does not guarantee migration however, but it is statistically more likely to occur.}  As the packing fraction increases, so does the number of particles which interact closely with their neighbors {\color{black} resulting in larger migratory excursions}. For the $\phi_B={\color{black}0.06}$ experiments, there are few particles with {\color{black}$\langle d_{\mathrm{min}} \rangle/(2a) < 0.2$} compared to the $\phi_B = {\color{black}0.10}$ or $\phi_B = {\color{black}0.12}$ experiments and consequently little migration towards the channel center, which is corroborated in the concentration profile results.  We observe no appreciable difference when the heat maps are generated using only data from before or after steady state is reached. However, the heat map results do exhibit some spatial dependence which is explored in Appendix B.  Example videos of particle behavior at specific points on the heat map are provided in the supplementary materials. {\color{black}Heat maps are also presented for the corresponding FCM simulations in figure \ref{fig:heatmap_simulation}. Since the simulations feature steady rather than periodic flow, the particle migrations are measured in 12 strain unit increments so that the average particle $z$-displacement between snapshots is the same as in the experiments. Like in the experiments, particles with more interactions are more likely to migrate and the particle proximity tends to increase with packing fraction.}

\begin{figure}
    \centerline{\includegraphics[width=\textwidth]{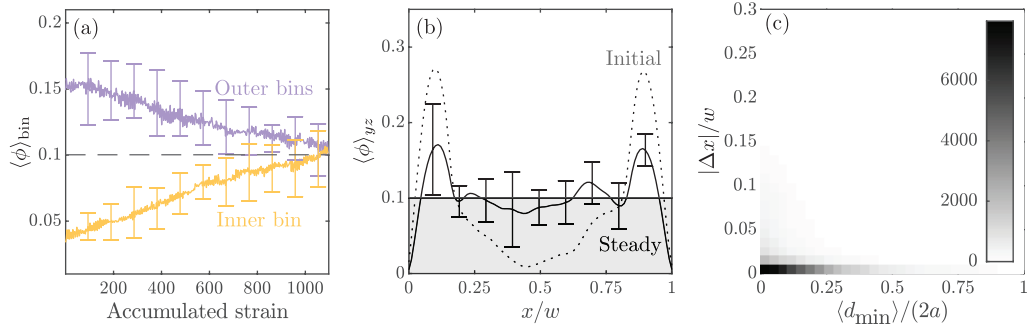}}
    \caption{Experiments are performed at $\phi_B={\color{black}0.10}$ and $\gamma=1$ and the (a) bin concentration evolution, (b) concentration profiles and (c) migration heat map are plotted. The initial concentration profile is averaged over the first 20 frames ($\gamma_a < 2.74$) and the steady profile over frames 7000 to 7400 ($\gamma_a = [960, 1016]$). {\color{black}The shaded region in (b) indicates the measured bulk packing fraction $\phi_B$.}}
    \label{fig:gamma_1}
\end{figure}

\subsection{{\color{black}Influence of strain amplitude}}

Experiments are also performed at $\phi_B={\color{black}0.10}$ {\color{black}($\phi_B'=0.11$)} and $\gamma=1$ in order to observe the influence of strain amplitude on the migration dynamics. The average particle concentrations in the center and outer quarter-width bins are plotted against the accumulated strain in figure \ref{fig:gamma_1}(a) up to an accumulated strain of 1100. Despite the large accumulated strain, the bin concentrations do not cross and the experiments exhibit quasi-reversible behavior where the bin concentrations change very slowly, potentially due to {\color{black} the locally high volume fraction at the walls.  Furthermore, the effects of residual buoyancy (due to distribution in particle density) also lead to irreversible behaviors that are more apparent on long timescales.} The initial and steady state concentration profiles in figure \ref{fig:gamma_1}(b) feature a similar lack of migration which is consistent with observations in the literature \citep{pham2016, guasto2010} of a critical strain amplitude below which the particles can subtly reorganize into a configuration that is essentially reversible. For the migration heat map in figure \ref{fig:gamma_1}(c), the peak intensity occurs near {\color{black}$\langle d_{\mathrm{min}} \rangle/(2a) = 0$} yet almost no migration occurs. Since the suspension is quasi-reversible at this low strain amplitude, the particles remain near their initial tightly packed wall-loaded configuration. {\color{black}Measurements of the mean squared particle displacements after integral numbers of cycles yield values that are at least an order of magnitude smaller than those reported in figure \ref{fig:diffusion}(a). However, given the particle tracking resolution, we cannot make a statistically significant estimate for the effective diffusivity without substantially more data.}

\section{Conclusions}

In our experiments, suspensions of neutrally buoyant spherical particles at bulk volume fractions $\phi_B$ from {\color{black}0.06} to {\color{black}0.12} are subjected to oscillatory pipe flow in a confined rectangular channel and the resulting {\color{black}particle} migration is characterized through direct imaging. {\color{black}We study a channel of aspect ratio 3:1 ($x$:$y$) with particles that are just larger than half the channel height in $y$ to prevent complete overlap.}  {\color{black}While numerous qualitative comparisons to the existing state-of-the-art measurements have been made throughout, quantitative comparisons are far more challenging, as it is not straight forward to unambiguously map between highly disparate flow geometries.}  

In experiments performed at strain amplitude $\gamma=6$, the particles are observed to migrate preferentially {\color{black}in the spanwise $x$-direction} towards the channel center, with higher bulk volume fraction hastening the dynamics. These findings are consistent with results in the literature for other, unconfined flow geometries {\color{black} exhibiting shear-induced migration} {\color{black} \citep{snook15,lyon98a,guasto2010}, although the enhanced concentration at the center of the channel is less pronounced in our case}. Other measurements of the collective particle behavior such as local particle concentration and velocity are similarly consistent with previous results {\color{black} \citep{lyon98a,guasto2010,snook15}}. The {\color{black} relative} particle concentration at the channel center {\color{black} increases} as the bulk volume fraction is increased. Conversely, the velocity profile is increasingly blunted compared to the theoretical $\phi_B=0$ case as $\phi_B$ is increased. {\color{black}Measurements of particle trajectories show that individual particles exhibit anisotropic diffusive-like behaviors, with streamwise diffusivities about an order of magnitude larger than their spanwise counterparts, again consistent with prior work \citep{pine05,guasto2010}.}
Additional experiments at a lower strain amplitude $\gamma=1$ produce quasi-reversible particle behavior consistent with the existence of a critical strain amplitude below which shear migration does not occur \citep{pham2016,guasto2010}.  {\color{black}In the small cross-stream ($y$) cross-sectional dimension, particles rapidly migrate to the walls in all cases and reach a steady-state distribution more rapidly than in the $x$ direction.  This rapid segregation is also observed in companion Force Coupling Method simulations where the spheres are not confined in the $x$-direction.} 

{\color{black}Furthermore, other comparisons to the literature indicate that confinement plays an impactful role in our system. In particular, previous results indicate a lack of migration for low volume fractions around 10\% \citep{hampton1997migration,snook15}, whereas we find clear evidence for migratory behavior at these low densities.  Additionally, the time to steady state in our system is comparable to that of more dense (but less confined) suspensions \citep{snook15}.  Both of these observations suggest that confinement tends to enhance irreversibility.  Geometrical confinement restricts possible particle motions, and thus naturally provides a mechanism for increasing the number of particle contacts when sheared. Lastly, for the case of extreme confinement (i.e. in the small cross-stream direction of our channel), the wall layers observed in prior work \citep{snook15} entirely dominate the resultant distribution.  Future experiments that more continuously span from high to low levels of confinement using a single flow geometry may provide additional valuable insight into the role of confinement on non-uniformly sheared non-Brownian suspensions.}

{\color{black}In order to conduct the experimental measurements,} we also implement a new, single-camera 3D particle tracking method. 
The transparent particles are imaged above a speckle pattern and the magnitude of the resulting optical distortions are used to infer the out-of-plane particle positions with high accuracy. This method trades the high optical component cost of typical {\color{black}particle suspension} setups for computational complexity, utilizing a refraction model to predict the distortion of the speckle patterns and iterating the particle position 
to reproduce the observed distortion.

The 3D tracking method permits the characterization of not only collective, but also individual particle behavior, enabling direct observations of {\color{black} interparticle spacing} and trajectories over long times. Consequently, it is possible to build a quantitative connection between particle interactions and displacement over a cycle to explain the source of {\color{black} irreversibility}. {\color{black} In both experiments and simulation} it is observed that, as bulk volume fraction increases, a greater number of particles experience close interactions with their neighbors and these same particles are responsible for the strongest individual migration events. {\color{black} While similar measurements have been performed for very few numbers of particles in 2D \citep{pham2015}, this work represents the first realization of similar temporally resolved, high-resolution trajectories for suspensions of large numbers of particles in 3D.}

In addition to allowing measurements of otherwise inaccessible quantities in the mono-disperse case, our 3D tracking method has the advantage of convenient extension to the case of bi-disperse or poly-disperse suspensions as particle size can be easily determined during the optical reconstruction. Due to the presence of the channel walls which prevent particle overlap, the present experimental findings are also relevant to natural problems involving particle suspension in high confinement such as thin films \citep{hooshanginejad2019stability} or porous media \citep{mirbod2023}. Furthermore, as a consequence of the standard manufacturing methods, duct-like channels are extremely common in microfluidics.
It has previously been shown that inertial effects in such geometries lead to focusing of suspensions in certain regions of the channel cross-section and can be exploited for passive particle manipulation \citep{di2009inertial}.  Irreversible particle-particle interactions in duct-like geometries can also result in {\color{black}particle migration across streamlines}, even in the absence of inertia, {\color{black} with confinement hastening the process}.

\vspace{3mm}
{\small 
\noindent {\bf Supplementary material.} Supplementary videos are made available to the interested reader.\\
\noindent {\bf Funding.} AH acknowledges support from the U.S. Department of Energy, Advanced Scientific Computing Research program, under the Scalable, Efficient and Accelerated Causal Reasoning Operators, Graphs and Spikes for Earth and Embedded Systems (SEA-CROGS) project, FWP 80278. Pacific Northwest National Laboratory (PNNL) is a multi-program national laboratory operated for the U.S. Department of Energy (DOE) by Battelle Memorial Institute under Contract No. DE-AC05-76RL01830. FV acknowledges funding from the University of Granada through the Brown/CASA-UGR Research Collaboration Fund and MICINN PID2019-104883GB-I00 project (Spain). \\
\noindent {\color{black} {\bf Acknowledgements.} All authors would like to thank the referees for their constructive suggestions.} \\
\noindent {\bf Code and data accessibility.} 
Particle tracking code, associated documentation, and data sets can be found at \url{https://github.com/harrislab-brown/ParticleHeight}. \\
{\bf Declaration of interests.} The authors report no conflict of interest.}

{\color{black}
\section*{Appendix A. Uniform initial conditions}

\begin{figure}
    \centerline{\includegraphics[width=\textwidth]{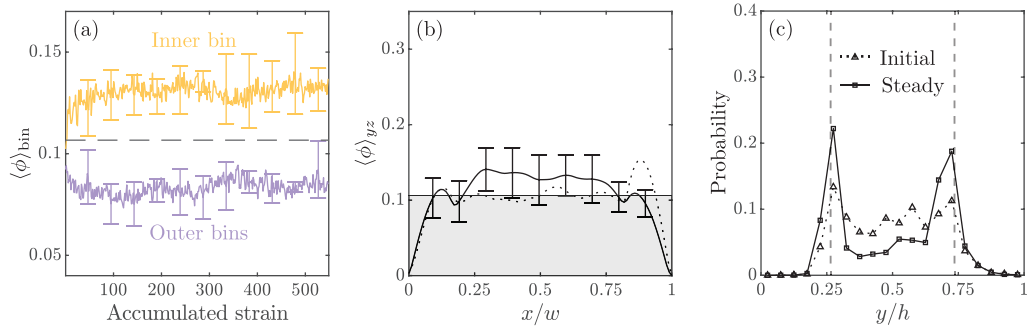}}
    \caption{{\color{black} Experiments are performed at $\phi_B=0.11$ and $\gamma=6$ with a homogeneous initial concentration yet the steady state of the suspension is {\color{black}similar to the case with} wall-loaded initial conditions. (a) The bin concentration evolution is plotted, with the error bars showing the standard deviation of 5 trials. (b) $x$-concentration initial (dotted) and steady (solid) profiles. The initial concentration profile is averaged over the first 20 frames ($\gamma_a < 2.74$) and the steady profile over frames 3500 to 3700 ($\gamma_a = [480, 508]$). {\color{black}The shaded region indicates the measured bulk packing fraction $\phi_B$.} (c) Initial (dotted) and steady (solid) distributions of particle centers in the $y$-direction. The likelihood that a given particle's center will lie in each region of the channel is plotted. The initial distribution is averaged over the first 20 frames ($\gamma_a < 2.74$) and the steady profile over frames 601 to 656 ($\gamma_a = [82, 90]$). The vertical dashed lines indicate a distance of one particle radius from the wall.}}
    \label{fig:uniform}
\end{figure}

In order to {\color{black}explore the sensitivity of the} experimental steady state particle distribution to the wall-loaded initial conditions, we also performed experiments at $\phi_B=0.11$ {\color{black}($\phi_B'=0.11$)} and $\gamma=6$ with an approximately homogeneous initial particle distribution. The resulting migration dynamics and particle distribution profiles across both channel dimensions are shown in figure \ref{fig:uniform}. Although the initial conditions here are notably different, we recover {\color{black} a similar} steady state as the experiments with a wall-loaded initial configuration, as can be seen by comparing with figures \ref{fig:narrow_migration_exp}(b) and \ref{fig:profiles}(a). An approximately equal number of particles start in the inner and outer bins in this case, with particles preferentially migrating to the inner bin as time progresses. Similarly, the bin concentrations recover a state close to the equilibrium values in figure \ref{fig:bin_evolution}(b).

\section*{Appendix B. Migration heat maps in inner and outer bins}

\begin{figure}
    \centerline{\includegraphics[width=\textwidth]{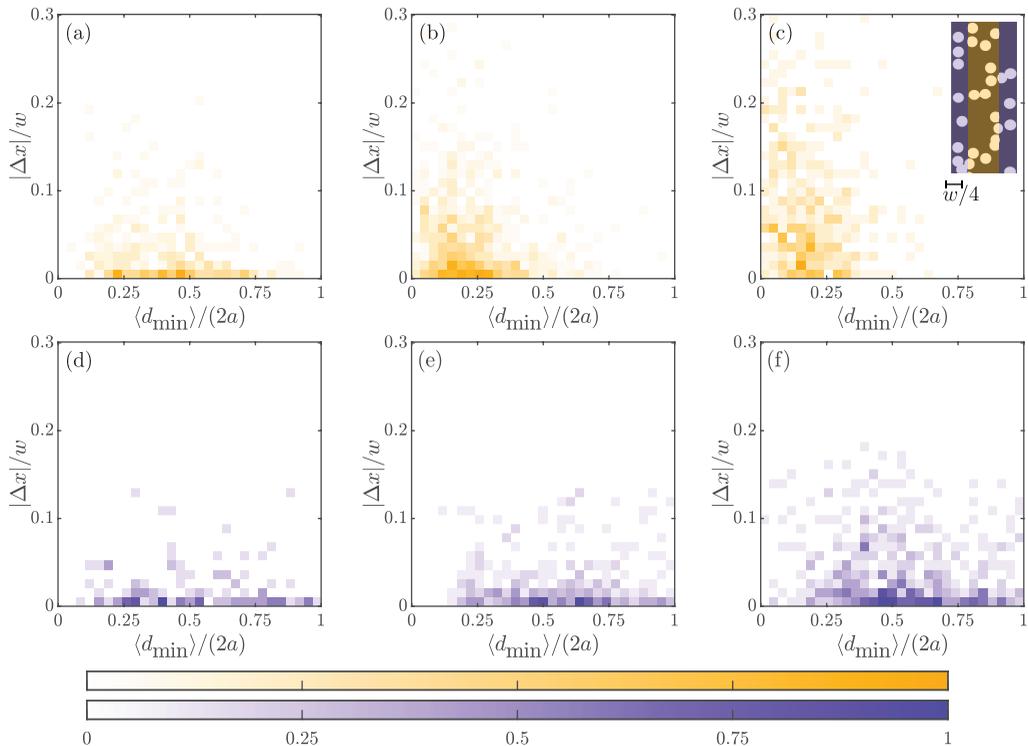}}
    \caption{{\color{black}Experimental heat maps after steady state of $x$-particle migration over one cycle versus average distance to nearest neighbor at $\gamma=6$ for (a) \& (d) $\phi_B={\color{black}0.06}$, (b) \& (e) $\phi_B={\color{black}0.10}$ and (c) \& (f) $\phi_B={\color{black}0.12}$. Only particles which start a cycle in the inner quarter-width bins (a) - (c) or outer quarter width bins (d) - (f) are considered. The color map is normalized by the maximum bin count which is 15, 20, 10, 7, 11, and 10, respectively.}}
    \label{fig:heatmap_bins}
\end{figure}

Heat maps of $x$-particle migration generated using only data from after steady state are shown in figure \ref{fig:heatmap_bins} in order to explore the spatial dependence of the particle behavior. Figure \ref{fig:heatmap_bins}(a) through (c) use only particle trajectories which start the cycle in the center quarter-width bins. Conversely, figures \ref{fig:heatmap_bins}(d) through (f) use only particle trajectories which start the cycle in the outer quarter-width bins. The inner bin migration heat maps closely resemble the full channel experimental heat maps in figure \ref{fig:heatmap} as well as the simulation heat maps in figure \ref{fig:heatmap_simulation}, the latter of which are fully unconfined in the spanwise ($x$) direction. On the other hand, the outer bin migration heat maps -- where the confinement from the channel walls plays a role -- are notably different. Although the average particle remains further from its neighbors in this region (since the average outer bin volume fraction is diminished once steady state is reached), even particles which interact closely never achieve migration events as large as those in the center bins, which is consistent with the existence of a persistent wall layer in which individual particles may become ``trapped'' for long times.  
}
\textcolor{black}{Particle wall layering has been shown in simulations of wall-bounded Couette and Poiseuille flow suspensions across a range of systems \citep{kulkarni2008suspension, Yeo2010, Yeo2011, kanehl2015hydrodynamic}, and similar wall layering has been seen in experiments \citep{snook15}. Physically, if two particles of equal size are in the wall layer and are involved in a collision, the force between the two particles is along the line of centers of the particles, which is parallel to the wall. Thus, there is no force to move the particles from the wall layer. Additionally, collisions with particles above the wall layer can serve to push particles in the wall layer closer to the wall. This results in a consistent and stable monolayer forming on the channel walls, with very few particles leaving over long times \citep{howard2018thesis}.}

\bibliographystyle{jfm} 
\bibliography{references}

\begin{thebibliography}{48}
\expandafter\ifx\csname natexlab\endcsname\relax\def\natexlab#1{#1}\fi
\def\au#1{#1} \def\ed#1{#1} \def\yr#1{#1}\def\at#1{#1}\def\jt#1{\textit{#1}} \def\bt#1{#1}\def\bvol#1{\textbf{#1}} \def\vol#1{#1} \def\pg#1{#1} \def\publ#1{#1}\def\arxiv#1{#1}\def\org#1{#1}\def\st#1{\textit{#1}}

\bibitem[Allan {\em et~al.\/}(2021)Allan, Caswell, Keim, van~der Wel \& Verweij]{trackpy}
{\sc \au{Allan, D.~B.}, \au{Caswell, T.}, \au{Keim, N.~C.}, \au{van~der Wel, C.~M.} \& \au{Verweij, R.~W.}} \yr{2021} soft-matter/trackpy: Trackpy v0.5.0.

\bibitem[Blanc {\em et~al.\/}(2011)Blanc, Peters \& Lemaire]{blanc2011}
{\sc \au{Blanc, F.}, \au{Peters, F.} \& \au{Lemaire, E.}} \yr{2011}  \at{Experimental signature of the pair trajectories of rough spheres in the shear-induced microstructure in noncolloidal suspensions}.  \jt{Physical Review Letters}  \bvol{107}~(20),  \pg{208302}.

\bibitem[Bradski(2000)]{opencv}
{\sc \au{Bradski, G.}} \yr{2000}  \at{{The OpenCV Library}}.  \jt{Dr. Dobb's Journal of Software Tools} .

\bibitem[Breedveld {\em et~al.\/}(2002)Breedveld, van~den Ende, Bosscher, Jongschaap \& Mellema]{breedveld2002}
{\sc \au{Breedveld, V.}, \au{van~den Ende, D.}, \au{Bosscher, M.}, \au{Jongschaap, R. J.~J.} \& \au{Mellema, J.}} \yr{2002}  \at{Measurement of the full shear-induced self-diffusion tensor of noncolloidal suspensions}.  \jt{The Journal of Chemical Physics}  \bvol{116}~(23),  \pg{10529--10535}.

\bibitem[Breedveld {\em et~al.\/}(2001)Breedveld, van~den Ende, Jongschaap \& Mellema]{breedveld2001}
{\sc \au{Breedveld, V.}, \au{van~den Ende, D.}, \au{Jongschaap, R.} \& \au{Mellema, J.}} \yr{2001}  \at{Shear-induced diffusion and rheology of noncolloidal suspensions: Time scales and particle displacements}.  \jt{The Journal of Chemical Physics}  \bvol{114}~(13),  \pg{5923--5936}.

\bibitem[Cheng(2008)]{cheng08}
{\sc \au{Cheng, N.~S.}} \yr{2008}  \at{Formula for viscosity of glycerol-water mixture}.  \jt{Industrial and Engineering Chemistry Research}  \bvol{47},  \pg{3285--3288}.

\bibitem[Corte {\em et~al.\/}(2008)Corte, Chaikin, Gollub \& Pine]{corte2008}
{\sc \au{Corte, L.}, \au{Chaikin, P.~M.}, \au{Gollub, J.~P.} \& \au{Pine, D.~J.}} \yr{2008}  \at{Random organization in periodically driven systems}.  \jt{Nature Physics}  \bvol{4}~(5),  \pg{420--424}.

\bibitem[Crocker \& Grier(1996)]{crocker96}
{\sc \au{Crocker, J.~C.} \& \au{Grier, D.~G.}} \yr{1996}  \at{Methods of digital video microscopy for colloidal studies}.  \jt{Journal of Colloid and Interface Science}  \bvol{179}~(1),  \pg{298--310}.

\bibitem[Di~Carlo(2009)]{di2009inertial}
{\sc \au{Di~Carlo, D.}} \yr{2009}  \at{Inertial microfluidics}.  \jt{Lab on a Chip}  \bvol{9}~(21),  \pg{3038--3046}.

\bibitem[Drazer {\em et~al.\/}(2002)Drazer, Koplik, Khusid \& Acrivos]{drazer2002}
{\sc \au{Drazer, G.}, \au{Koplik, J.}, \au{Khusid, B.} \& \au{Acrivos, A.}} \yr{2002}  \at{Deterministic and stochastic behaviour of non-{B}rownian spheres in sheared suspensions}.  \jt{Journal of Fluid Mechanics}  \bvol{460},  \pg{307--335}.

\bibitem[Guasto {\em et~al.\/}(2010)Guasto, Ross \& Gollub]{guasto2010}
{\sc \au{Guasto, J.~S.}, \au{Ross, A.~S.} \& \au{Gollub, J.~P.}} \yr{2010}  \at{Hydrodynamic irreversibility in particle suspensions with nonuniform strain}.  \jt{Physical Review E}  \bvol{81},  \pg{061401}.

\bibitem[Guazzelli \& Pouliquen(2018)]{guazzelli2018rheology}
{\sc \au{Guazzelli, E.} \& \au{Pouliquen, O.}} \yr{2018}  \at{Rheology of dense granular suspensions}.  \jt{Journal of Fluid Mechanics}  \bvol{852},  \pg{P1}.

\bibitem[Hampton {\em et~al.\/}(1997)Hampton, Mammoli, Graham, Tetlow \& Altobelli]{hampton1997migration}
{\sc \au{Hampton, R.~E.}, \au{Mammoli, A.~A.}, \au{Graham, A.~L.}, \au{Tetlow, N.} \& \au{Altobelli, S.A.}} \yr{1997}  \at{Migration of particles undergoing pressure-driven flow in a circular conduit}.  \jt{Journal of rheology}  \bvol{41}~(3),  \pg{621--640}.

\bibitem[Hecht(2012)]{hecht2012}
{\sc \au{Hecht, E.}} \yr{2012} {\em Optics\/}.  \publ{Pearson}.

\bibitem[Hooshanginejad {\em et~al.\/}(2019)Hooshanginejad, Druecke \& Lee]{hooshanginejad2019stability}
{\sc \au{Hooshanginejad, A.}, \au{Druecke, B.~C.} \& \au{Lee, S.}} \yr{2019}  \at{Stability analysis of a particle band on the fluid--fluid interface}.  \jt{Journal of Fluid Mechanics}  \bvol{869}.

\bibitem[Howard(2018)]{howard2018thesis}
{\sc \au{Howard, A.~A.}} \yr{2018}  \at{Numerical simulations to investigate particle dispersion in non-homogeneous suspension flows}. PhD thesis, Brown University.

\bibitem[Howard \& Maxey(2018)]{howard2018}
{\sc \au{Howard, A.~A.} \& \au{Maxey, M.~R.}} \yr{2018}  \at{Simulation study of particle clouds in oscillating shear flow}.  \jt{Journal of Fluid Mechanics}  \bvol{852},  \pg{484--506}.

\bibitem[Ingber {\em et~al.\/}(2006)Ingber, Mammoli, Vorobieff, McCollam \& Graham]{ingber2006}
{\sc \au{Ingber, M.~S.}, \au{Mammoli, A.~A.}, \au{Vorobieff, P.}, \au{McCollam, T.} \& \au{Graham, A.~L.}} \yr{2006}  \at{Experimental and numerical analysis of irreversibilities among particles suspended in a {C}ouette device}.  \jt{Journal of Rheology}  \bvol{50}~(2),  \pg{99--114}.

\bibitem[Johnson(2021)]{nlopt}
{\sc \au{Johnson, S.~G.}} \yr{2021} The {NLopt} nonlinear-optimization package.

\bibitem[Kanehl \& Stark(2015)]{kanehl2015hydrodynamic}
{\sc \au{Kanehl, P.} \& \au{Stark, H.}} \yr{2015}  \at{Hydrodynamic segregation in a bidisperse colloidal suspension in microchannel flow: A theoretical study}.  \jt{The Journal of chemical physics}  \bvol{142}~(21).

\bibitem[Kilbride {\em et~al.\/}(2023)Kilbride, Fagg, Ouali \& Fairhurst]{kilbride2023pattern}
{\sc \au{Kilbride, J.~J.}, \au{Fagg, K.~E.}, \au{Ouali, F.~F.} \& \au{Fairhurst, D.~J.}} \yr{2023}  \at{Pattern-distortion technique: Using liquid-lens magnification to extract volumes of individual droplets or bubbles within evaporating two-dimensional arrays}.  \jt{Physical Review Applied}  \bvol{19}~(4),  \pg{044030}.

\bibitem[Kulkarni \& Morris(2008)]{kulkarni2008suspension}
{\sc \au{Kulkarni, P.~M.} \& \au{Morris, J.~F.}} \yr{2008}  \at{Suspension properties at finite reynolds number from simulated shear flow}.  \jt{Physics of Fluids}  \bvol{20}~(4).

\bibitem[Lyon \& Leal(1998{\natexlab{{\em a\/}}})]{lyon98a}
{\sc \au{Lyon, M.~K.} \& \au{Leal, L.~G.}} \yr{1998{\natexlab{{\em a\/}}}}  \at{An experimental study of the motion of concentrated suspensions in two-dimensional channel flow. part 1. monodisperse systems}.  \jt{Journal of Fluid Mechanics}  \bvol{363},  \pg{25--56}.

\bibitem[Lyon \& Leal(1998{\natexlab{{\em b\/}}})]{lyon98b}
{\sc \au{Lyon, M.~K.} \& \au{Leal, L.~G.}} \yr{1998{\natexlab{{\em b\/}}}}  \at{An experimental study of the motion of concentrated suspensions in two-dimensional channel flow. part 2. bidisperse systems}.  \jt{Journal of Fluid Mechanics}  \bvol{363},  \pg{57--77}.

\bibitem[Marnoto \& Hashmi(2023)]{marnoto2023application}
{\sc \au{Marnoto, A.} \& \au{Hashmi, S.~M.}} \yr{2023}  \at{Application of droplet migration scaling behavior to microchannel flow measurements}.  \jt{Soft Matter} .

\bibitem[Maxey(2017)]{maxey2017}
{\sc \au{Maxey, M.}} \yr{2017}  \at{Simulation methods for particulate flows and concentrated suspensions}.  \jt{Annual Review of Fluid Mechanics}  \bvol{49},  \pg{171--193}.

\bibitem[Metzger \& Butler(2010)]{metzger2010}
{\sc \au{Metzger, B.} \& \au{Butler, J.~E.}} \yr{2010}  \at{Irreversibility and chaos: Role of long-range hydrodynamic interactions in sheared suspensions}.  \jt{Physical Review E}  \bvol{82}~(5),  \pg{051406}.

\bibitem[Metzger \& Butler(2012)]{metzger2012}
{\sc \au{Metzger, B.} \& \au{Butler, J.~E.}} \yr{2012}  \at{Clouds of particles in a periodic shear flow}.  \jt{Physics of Fluids}  \bvol{24}~(2),  \pg{021703}.

\bibitem[Metzger {\em et~al.\/}(2013)Metzger, Pham \& Butler]{metzger2013}
{\sc \au{Metzger, B.}, \au{Pham, P.} \& \au{Butler, J.~E.}} \yr{2013}  \at{Irreversibility and chaos: Role of lubrication interactions in sheared suspensions}.  \jt{Physical Review E}  \bvol{87}~(5),  \pg{052304}.

\bibitem[Mirbod \& Shapley(2023)]{mirbod2023}
{\sc \au{Mirbod, P.} \& \au{Shapley, N.~C.}} \yr{2023}  \at{Particle migration of suspensions in a pressure-driven flow over and through a porous structure}.  \jt{Journal of Rheology}  \bvol{67}~(2),  \pg{417--432}.

\bibitem[Moisy {\em et~al.\/}(2009)Moisy, Rabaud \& Salsac]{moisy2009synthetic}
{\sc \au{Moisy, F.}, \au{Rabaud, M.} \& \au{Salsac, K.}} \yr{2009}  \at{A synthetic schlieren method for the measurement of the topography of a liquid interface}.  \jt{Experiments in Fluids}  \bvol{46}~(6),  \pg{1021--1036}.

\bibitem[More \& Ardekani(2020)]{more2020constitutive}
{\sc \au{More, R.~V.} \& \au{Ardekani, A.~M.}} \yr{2020}  \at{A constitutive model for sheared dense suspensions of rough particles}.  \jt{Journal of Rheology}  \bvol{64}~(5),  \pg{1107--1120}.

\bibitem[Morris(2001)]{morris2001anomalous}
{\sc \au{Morris, J.~F.}} \yr{2001}  \at{Anomalous migration in simulated oscillatory pressure-driven flow of a concentrated suspension}.  \jt{Physics of Fluids}  \bvol{13}~(9),  \pg{2457--2462}.

\bibitem[Morris \& Boulay(1999)]{morris1999curvilinear}
{\sc \au{Morris, J.~F.} \& \au{Boulay, F.}} \yr{1999}  \at{Curvilinear flows of noncolloidal suspensions: The role of normal stresses}.  \jt{Journal of rheology}  \bvol{43}~(5),  \pg{1213--1237}.

\bibitem[Nelder \& Mead(1965)]{nelder65}
{\sc \au{Nelder, J.~A.} \& \au{Mead, R.}} \yr{1965}  \at{{A Simplex Method for Function Minimization}}.  \jt{The Computer Journal}  \bvol{7}~(4),  \pg{308--313}.

\bibitem[Nott \& Brady(1994)]{nott1994pressure}
{\sc \au{Nott, P.~R.} \& \au{Brady, J.~F.}} \yr{1994}  \at{Pressure-driven flow of suspensions: simulation and theory}.  \jt{Journal of Fluid Mechanics}  \bvol{275},  \pg{157--199}.

\bibitem[Olufsen {\em et~al.\/}(2019)Olufsen, Andersen \& Fagerholt]{olufsen19}
{\sc \au{Olufsen, S.~N.}, \au{Andersen, M.~E.} \& \au{Fagerholt, E.}} \yr{2019}  \at{µdic: An open-source toolkit for digital image correlation}.  \jt{SoftwareX}  \bvol{11}.

\bibitem[Pham {\em et~al.\/}(2016)Pham, Butler \& Metzger]{pham2016}
{\sc \au{Pham, P.}, \au{Butler, J.~E.} \& \au{Metzger, B.}} \yr{2016}  \at{Origin of critical strain amplitude in periodically sheared suspensions}.  \jt{Physical Review Fluids}  \bvol{1}~(2),  \pg{022201}.

\bibitem[Pham {\em et~al.\/}(2015)Pham, Metzger \& Butler]{pham2015}
{\sc \au{Pham, P.}, \au{Metzger, B.} \& \au{Butler, J.~E.}} \yr{2015}  \at{Particle dispersion in sheared suspensions: Crucial role of solid-solid contacts}.  \jt{Physics of Fluids}  \bvol{27}~(5),  \pg{051701}.

\bibitem[Pine {\em et~al.\/}(2005)Pine, Gollub, Brady \& Leshansky]{pine05}
{\sc \au{Pine, D.~J.}, \au{Gollub, J.~P.}, \au{Brady, J.~F.} \& \au{Leshansky, A.~M.}} \yr{2005}  \at{Chaos and threshold for irreversibility in sheared suspensions}.  \jt{Nature}  \bvol{438}~(7070),  \pg{997–1000}.

\bibitem[Rampall {\em et~al.\/}(1997)Rampall, Smart \& Leighton]{rampall1997}
{\sc \au{Rampall, I.}, \au{Smart, J.~R.} \& \au{Leighton, D.~T.}} \yr{1997}  \at{The influence of surface roughness on the particle-pair distribution function of dilute suspensions of non-colloidal spheres in simple shear flow}.  \jt{Journal of Fluid Mechanics}  \bvol{339},  \pg{1--24}.

\bibitem[Rintoul \& Torquato(1996)]{rintoul96}
{\sc \au{Rintoul, M.~D.} \& \au{Torquato, S.}} \yr{1996}  \at{Computer simulations of dense hard‐sphere systems}.  \jt{The Journal of Chemical Physics}  \bvol{105}~(20),  \pg{9258--9265},  \arxiv{arXiv: https://pubs.aip.org/aip/jcp/article-pdf/105/20/9258/9438512/9258\_1\_online.pdf}.

\bibitem[Snook {\em et~al.\/}(2015)Snook, Butler \& Guazelli]{snook15}
{\sc \au{Snook, B.}, \au{Butler, J.~E.} \& \au{Guazelli, E.}} \yr{2015}  \at{Dynamics of shear-induced migration of spherical particles in oscillatory pipe flow}.  \jt{Journal of Fluid Mechanics}  \bvol{786},  \pg{128--153}.

\bibitem[Vincent(1993)]{vincent93}
{\sc \au{Vincent, L.}} \yr{1993}  \at{Morphological grayscale reconstruction in image analysis: applications and efficient algorithms}.  \jt{IEEE Transactions on Image Processing}  \bvol{2}~(2),  \pg{176--201}.

\bibitem[Yeo \& Maxey(2010)]{Yeo2010}
{\sc \au{Yeo, K.} \& \au{Maxey, M.~R.}} \yr{2010}  \at{Simulation of concentrated suspensions using the force-coupling method}.  \jt{J. Comput. Phys.}  \bvol{229},  \pg{2401--2421}.

\bibitem[Yeo \& Maxey(2011)]{Yeo2011}
{\sc \au{Yeo, K.} \& \au{Maxey, M.~R.}} \yr{2011}  \at{Numerical simulations of concentrated suspensions of monodisperse particles in a {Poiseuille} flow}.  \jt{J. Fluid Mech.}  \bvol{682},  \pg{491--518}.

\bibitem[Zhang {\em et~al.\/}(2023)Zhang, Pham, Metzger, Kopelevich \& Butler]{zhang2023effect}
{\sc \au{Zhang, H.}, \au{Pham, P.}, \au{Metzger, B.}, \au{Kopelevich, D.~I.} \& \au{Butler, J.~E.}} \yr{2023}  \at{Effect of particle roughness on shear-induced diffusion}.  \jt{Physical Review Fluids}  \bvol{8}~(6),  \pg{064303}.

\bibitem[Zivkovic(2004)]{zivkovic04}
{\sc \au{Zivkovic, Z.}} \yr{2004} Improved adaptive {G}aussian mixture model for background subtraction.  \bt{In {\em Proceedings of the 17th International Conference on Pattern Recognition\/}},  \st{ICPR},  \vol{vol.~2},  \pg{pp. 28--31}.

\end{thebibliography}

\end{document}